\documentclass[12pt]{iopart}

\usepackage{iopams}  
\usepackage{epsfig}
\usepackage{graphicx}
\usepackage{cite}
\def\eq#1{{Eq.~(\ref{#1})}}
\def\beq{\begin{equation}}
\def\eeq{\end{equation}}
\def\bea{\begin{eqnarray}}
\def\eea{\end{eqnarray}}

\def\fig#1{{Fig.~\ref{#1}}}
\def\ud{\underline}

\def\iq{\!=\!}
\begin{document}

\title[]{Particle
  multiplicities in nuclear collisions at the LHC from saturation physics}

\author{Javier L. Albacete}

\address{The Ohio State University. Department of Physics. \\
191 W. Woodruff Ave, Columbus OH-43210, USA}
\ead{albacete@mps.ohio-state.edu}
\begin{abstract}
  The inclusion of running coupling effects in the BK-JIMWLK evolution
  equations considerably reduces the energy dependence of the
  saturation scale with respect to previous estimates based on fits to
  HERA data. We discuss how such slowdown affects the expectations for
  particle multiplicities in Pb-Pb collisions at LHC energies. Our
  prediction is based on the use of $k_t$-factorization and on the use
  of unintegrated gluon distributions taken from the numerical
  solutions of the BK equation with running coupling. We obtain a
  central value $dN^{ch}/d\eta(\sqrt{s}=5.5\,
  \mbox{TeV})\vert_{\eta=0}\approx 1390$.
\end{abstract}
\pacs{21.65 Qr, 12.38 Mh}

The experimental results from RHIC \cite{rhic} strongly suggest that heavy ion
collisions at high energies probe QCD in the non-linear regime
characterized by strong coherent fields and gluon saturation, the
Color Glass Condensate (CGC) \cite{cgc}. Thus, many bulk features of
multiparticle production in RHIC collisions, such as the energy,
rapidity and centrality dependence of multiparticle production, are
succesfully described by models based in CGC physics \cite{kln,asw}.
With collision energies of up to $5.5$ TeV, the upcoming program in
lead-lead collisions at the CERN Large Hadron Collider (LHC) is expected to
provide confirmation for the tentative  
conclusions reached at RHIC and to discriminate between the different
physical mechanisms proposed to explain particle production in high
energy nuclear reactions (a review of predictions for heavy ion
collisions at the LHC based on alternative approaches can be
found in \cite{cernlhc}). 

The phenomenological models in \cite{kln,asw} rely on the
assumption that 
the saturation  scale $Q_{sA}$ that governs the onset of non-linear
effects in the wave function 
of the colliding nuclei is perturbatively large $\sim\!1\!$ GeV at the highest
RHIC energies. Next, gluon production is calculated via the
convolution of the nuclear unintegrated gluon distributions (ugd's)
according to $k_t$-factorization \cite{glr}. Under the additional
assumption of local parton-hadron duality, the pseudorapidity density of charged
particles produced in a nucleus-nucleus collisions can be written as follows
\begin{equation}\label{fact}
\!\!\!\!\!\!\!\!\!\!\!\!\!\!\!\!\!\frac{dN_{ch}}{dy\, d^2b}=C\frac{4\pi
  N_c}{N_c^2-1}\int\frac{d^2p_t}{p_t^2}\int d^2k_t\,
\alpha_s(Q)\,\varphi\left(x_1,\frac{\vert\ud{k_t}+\ud{p_t}\vert}{2}\right)
\varphi\left(x_2,\frac{\vert\ud{k_t}-\ud{p_t}\vert}{2}\right), 
\label{ktfact}
\end{equation}
where $\varphi$ denote the ugd's of projectile and target
respectively, $p_t$ and $y$ are the transverse momentum and rapidity
of the produced particle, $x_{1,2}\iq(p_t/\sqrt{s})\,e^{\pm y}$,
$Q\iq0.5\max\left\{\vert p_t\pm k_t\vert \right\}$ and $b$ the impact
parameter of the collision. The lack of impact parameter integration
in this calculation and the gluon to charged hadron ratio are
accounted for by the constant $C$, which sets the normalization.

The second basic ingredient of CGC models is the one of saturation of
the nuclear udg's entering \eq{fact} Using a relatively simple ansatz
for the nuclear ugd's, and for symmetric collisions, the midrapidity
multiplicity rises proportional to the nuclear saturation scale, which is
assumed to grow as a power of the collision energy, $\sqrt{s}$:
\begin{equation}
\left.\frac{dN_{ch}}{dy\, d^2b}\right\vert_{y=0}\propto Q_{sA}^2\approx
const\cdot \sqrt{s}^{\lambda}.
\end{equation}
So far, the energy dependence of the saturation scale has been
adjusted to the empirical value extracted from fits to DIS HERA data
of saturation-based models\cite{gbw}, which yields $\lambda\approx
0.288$. This has been largely motivated by the inability of the
leading-log BK-JIMWLK evolution equations to reproduce experimental
data. Here discuss how the recent advances in the calculation of
running coupling corrections to the BK-JIMWLK evolution equations
\cite{bkrc} allow to compute the energy evolution of the ugd's from
first principles, getting a good agreement with experimental data.

\begin{figure}[ht]
\begin{center}
\includegraphics[height=7cm]{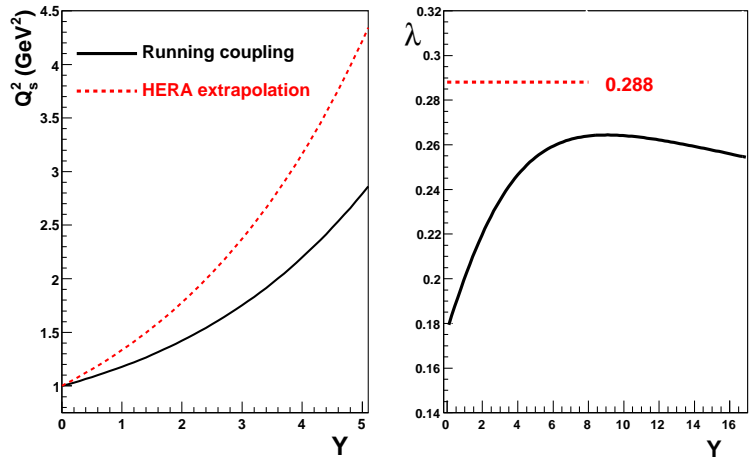}
\end{center}
\caption{ Left: Saturation scale as a function of rapidity
  corresponding to solutions of \eq{frs} (solid line) and setting
  $\lambda=0.288$ (dashed line). Right: Evolution speed,
  $\lambda=\frac{d\ln Q_s^2(Y)}{dY}$, corresponding to the plot in the
  left. In both cases the initial saturation scale is  $Q_{0}= 1$ GeV.
}\label{fig1}
\end{figure}

We start by solving the non-linear small-$x$ evolution equation
for the dipole-nucleus scattering 
matrix, $S(Y,r)$, including running coupling corrections
\cite{bkrc,Albacete:2007yr,Albacete:2007sm}:   
\begin{equation}\label{frs}
  \frac{\partial S(Y,r)}{\partial Y} \, = \,
  \mathcal{R}\left[S\right]-\mathcal{S}\left[S\right]\,,
\end{equation}
where $r$ is the dipole size and $Y=\ln(x_{0}/x)$.  Explicit
expressions for the kernel terms in \eq{frs} as well as a detailed
explanation of the numerical method used to solve \eq{frs} are given
in \cite{Albacete:2007yr}.  The initial conditions for the evolution
are taken from the McLerran-Venugopalan model \cite{mv}. We classify
them according to their initial value of the saturation scale,
$Q_0$. Importantly, the saturation scale extracted from solutions of
\eq{frs} grows with energy significantly more slowly than the value
extracted from empirical parametrizations of DIS data, as shown in
\fig{fig1}. On the other hand, the evolution speed
$\lambda=d\ln\,Q_s^2/dY$ turns out to be a function of $Y$, rather than
a constant.
%%%%%%%%%%%%%%%%%%
%%%%%%%%%%%%%%%%%

We now use the exact solutions of \eq{frs} to calculate particle
multiplicities according to \eq{ktfact}. Thus, the nuclear
udg's  are now given in terms of the dipole scattering
matrix evolved according to \eq{frs}:
\begin{equation}
   \varphi(Y,k) 
   = \int {d^2r\over 2\pi\, r^2}\exp\{i\, {\ud r}\cdot {\ud
     k}\}\,(1-S(Y,r))\, , 
   \label{phi}
\end{equation}
The relation between the evolution variable in \eq{frs} and
Feynman-$x$ of the produced particle is taken to be
$Y\!=\!\ln(0.05/x_{1,2})+\Delta Y_{ev}$.  Since the relevant values of
Bjorken-$x$ probed at mid-rapidities and $\sqrt{s_{NN}}\iq130$ GeV at
RHIC are estimated to be $\sim0.1\div0.01$, the free parameter $\Delta
Y_{ev}$ controls the extent of evolution undergone by the nuclear
gluon densities resulting of \eq{frs} prior to comparison with RHIC
data.  Similar to \cite{kln}, large-$x$ effects have been modelled by
replacing $\varphi(x,k)\rightarrow \varphi(x,k)(1-x)^4$.  The running
of the strong coupling in \eq{ktfact}, evaluated according to the one loop QCD
expression, is regularized in the infrared by freezing it to a
constant value $\alpha_{fr}\iq1$ at small momenta.  Finally, in order
to compare \eq{ktfact} with experimental data it is necessary to
correct the difference between rapidity, $y$, and the experimentally
measured pseudo-rapidity, $\eta$. This is achieved by introducing an
average hadron mass, $m$. Remarkably, the optimal value found in
comparison with data, $m\sim0.25$ GeV is in good quantitative
agreement with the hadrochemical composition of particle production at
RHIC.

\begin{figure}[ht]
\begin{center}
\includegraphics[height=7.5cm]{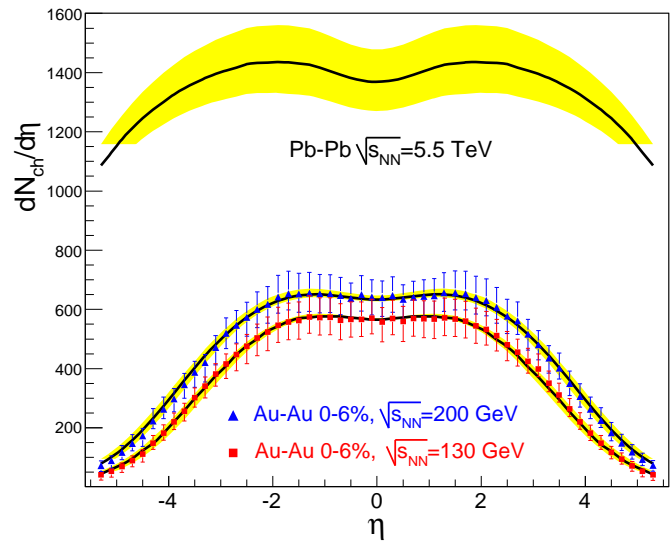}
\end{center}
\caption{Pseudo-rapidity density of charged particles produced in Au-Au 0-6\%
  central collisions at $\sqrt{s_{NN}}=130$ and 200 GeV and for Pb-Pb
  central collisions at $\sqrt{s_{NN}}=5.5$ TeV. Data taken from \cite{rhic}.
  The upper, central (solid
  lines) and lower limits of the theoretical uncertainty
  band correspond to ($Q_{0}\!=\!1$ 
  GeV, $\Delta Y\!=\!1$), ($Q_{0}\!=\!0.75$
  GeV, $\Delta Y\!=\!3$) and ($Q_{0}\!=\!1.25$ 
  GeV, $\Delta Y\!=\!0.5$) respectively, with $m\!=\!0.25$ GeV in all
  cases. 
}\label{fig2}
\end{figure}

With this set up we find a remarkably good agreement with the
pseudo-rapidity densities of charged particles measured in $0\!-\!6\%$
central Au+Au collisions at collision energies $\sqrt{s_{NN}}\iq130$
and 200 GeV, as shown in \fig{fig2}. The comparison with data
\cite{rhic} constrains the free parameters of the calculation to the
ranges: $Q_0\!\sim\!0.75\div1.25$ GeV, $m\!\sim\!0.25$ GeV and
$3\!\gtrsim\!\Delta Y_{ev}\!\gtrsim \!0.5$. Finally, the normalization
constant is $C\sim\mathcal{O}(1)$ in all cases. The smallness of $\Delta
Y_{ev}$ indicates that the nuclear udg's probed at RHIC are in the
{\it pre-asymptotic} regime \cite{aamsw}. With all the free
parameters now constrained by comparison to RHIC data, the extrapolation
to central Pb-Pb collisions at LHC energies, $\sqrt{s}=5.5$ TeV per
nucleon, is straightforward and completely driven by the
non-linear small-x dynamics. We get
\begin{equation}
  \label{pred}
 \left.\frac{dN^{Pb-Pb}_{ch}}{d\eta}(\sqrt{s_{NN}}\iq5.5\,\mbox{TeV},\eta=0)
\right.\sim1290\div1480\,,     
\end{equation}
with a central value corresponding to the best fits to RHIC data $\sim
1390$. These values are significantly smaller than those of other
saturation based calculations \cite{kln,asw,ekrt}, $\sim 1700\div2500$, and
compatible with the ones based on studies of the fragmentation region
\cite{gsv}, which yield $\approx 1000\div1400$.

%%%%%%%%%%%%%%%%%%%%%%%%%%%%%%%%%%%%%%%%%%%%%%%%%%%%%%%%%%%%%%%%%%%%%
%%%%%%%%%%%%%%%%%%%%%%%%%%%%%%%%%%%%%%%%%%%%%%%%%%%%%%%%%%%%%%%%%%%

This research is sponsored in part by the U.S. Department of Energy
under Grant No. DE-FG02-05ER41377 and by an allocation of computing
time from the Ohio Supercomputer Center.

\end{document}